\documentclass[pdflatex,sn-mathphys-num]{sn-jnl}

\usepackage{graphicx}%
\usepackage{multirow}%
\usepackage{amsmath,amssymb,amsfonts}%
\usepackage{amsthm}%
\usepackage{mathrsfs}%
\usepackage[title]{appendix}%
\usepackage{xcolor}%
\usepackage{textcomp}%
\usepackage{manyfoot}%
\usepackage{booktabs}%
\usepackage{algorithm}%
\usepackage{algorithmicx}%
\usepackage{algpseudocode}%
\usepackage{listings}%
\usepackage{soul}
\usepackage[capitalise]{cleveref}

\theoremstyle{thmstyleone}%
\theoremstyle{thmstyletwo}%

\theoremstyle{thmstylethree}%

\definecolor{internationalorange}{rgb}{1.0, 0.31, 0.0}

\begin{document}

\title[Article Title]{Polarization and Integration in Global AI Research}


\author[1,2]{\fnm{Luca} \sur{Gallo}}\email{lga@sodas.ku.net}

\author[3,4]{\fnm{Riccardo} \sur{Di Clemente}}\email{riccardo.diclemente@nulondon.ac.uk}

\author*[1,5,6]{\fnm{Balázs} \sur{Lengyel}}\email{lengyel.balazs@krtk.elte.hu}

\affil[1]{\small
\orgdiv{ANETI Lab, Corvinus Institute for Advanced Studies},
\orgname{Corvinus University of Budapest},
\orgaddress{\street{Közraktár u. 4-6}, \city{Budapest},
\postcode{1093}, \country{Hungary}}}

\affil[2]{\small
\orgdiv{Center for Social Data Science (SODAS)}, \orgname{University of Copenhagen}, \orgaddress{\street{Øster Farimagsgade 5A}, \city{Copenhagen}, \postcode{1353}, \country{Denmark}}}

\affil[3]{\small
\orgdiv{Complex Connections Lab, Network Science Institute}, \orgname{Northeastern University London}, \orgaddress{\city{London}, \postcode{E1W 1LP}, \country{United Kingdom}}}

\affil[4]{\small
\orgdiv{ISI Foundation}, \orgaddress{\street{Via Chisola 5}, \city{Turin}, \postcode{10126}, \country{Italy}}}

\affil[5]{\small
\orgdiv{MTA–ELTE ``Momentum'' Agglomeration, Networks, and Innovation Research Group}, \orgname{ELTE Centre for Economic and Regional Studies}, \orgaddress{\street{Tóth Kálmán u. 4}, \city{Budapest}, \postcode{1097}, \country{Hungary}}}

\affil[6]{\small
\orgdiv{Institute of Data Analytics and Information Systems}, \orgname{Corvinus University of Budapest}, \orgaddress{\street{Közraktár u. 4-6}, \city{Budapest}, \postcode{1093}, \country{Hungary}}}


\abstract{The AI race amplifies security risks and international tensions. While the US restricts mobility and knowledge flows, challenges regulatory efforts to protect its advantage, China leads initiatives of global governance. Both strategies depend on cross-country relationships in AI innovation; yet, how this system evolves is unclear. Here, we measure the processes of polarization and integration in the global AI research over three decades by using large-scale data of scientific publications. Comparing cross-country collaboration and citation links to their random realizations, we find that the US and China have long diverged in both dimensions, forming two poles around which global AI research increasingly revolves. While the United Kingdom and Germany have integrated exclusively with the US, many European countries have converged with both poles. Developing and further developed countries, however, only integrate with China, signaling its expanding influence over the international AI research landscape. Our results inform national science policies and efforts toward global AI regulations.}




\maketitle

\section*{Introduction}\label{sec:intro}

AI increasingly influences strategic areas of development like health \cite{he2019practical}, renewable energy \cite{ahmad2021artificial}, and scientific progress in general \cite{gao2024quantifying}; while it's growing strategic importance has made it a focal point of geopolitical competition. 
Western economies face a mounting challenge of loosing their innovation leadership as China's massive R$\&$D investments in future technologies accelerate \cite{wu2020towards,autor2025we, zhou2006emergence, xie2014china}. 
The dominant view is that China has vastly benefited from links to Western innovation hubs, which have helped the acceleration of its technological advancement \cite{saxenian2006new}. 
In response, the US has decided to cool research relationships with China and even investigated Chinese researchers working in the US to decrease foreign interference and theft of intellectual property within the heavily criticized China Initiative between 2018 and 2022 \cite{silver2020us,jia2024impact, kivelson2023high}.

Aiming for dominance in global AI production, the United States and China are pursuing fundamentally different regulatory strategies \cite{hutson2023rules}. 
The US federal government has loosened the already relatively weak regulations in 2025, potentially also affecting European regulatory efforts that have been significantly delayed recently. 
Meanwhile, China aims to lead the global governance initiative by establishing the World Artificial Intelligence Cooperation Organization (WAICO) to coordinate international AI regulation \cite{gibney2025china}.


Despite the fierce competition between the US and China and their cooling relationship, many argue that AI research needs stronger integration, as cross-border collaboration produces high-quality science \cite{chessa2013europe, alshebli2024china, schmallenbach2024global}.
Integrated innovation systems can also mitigate the risks of fragmented AI regulatory frameworks \cite{cihon2020fragmentation, schmitt2022mapping}. 
For example, the EU's new AI strategy for science will be introduced for the whole European Research Area (ERA) that contains not only the 27 EU member states but also 16 associated countries that is a clear signal that integration of the science system can extend AI regulation beyond political borders \cite{EC_COM_2025_724_biblatex}. 
The rivalry between the US and China unfolds in a scientific landscape more internationally interconnected than ever before.
International collaborations have grown rapidly for half a century, across multiple scientific fields \cite{luukkonen1992understanding,wagner2005network,coccia2016evolution,wagner2017growth}, reshaping the geography of science from US-European dominance toward an integrated, multipolar system in which China and other countries from the Global South emerge as novel scientific hubs\cite{gui2019globalization,ramirez2026transformation}.
Many speculate that the recent US restrictions could push China toward deeper partnerships with the EU and other countries \cite{churchill2018china}.   

However, how the structure of international relationships in AI research has evolved remains poorly understood.
Examining these dynamics can reveal whether the global AI innovation system is polarizing into competing spheres of influence or whether certain countries act as bridges of international research across the poles.

In this paper, we map the evolution of the complex network of international collaborations and knowledge flows in the global AI research over three decades, a period in which China has caught up to the global frontier in terms of co-authorships and citations. 
Rather than focusing on specific bilateral relationships, 
we analyze the entire global network to trace how the interaction patterns of China, the United States, and key European countries have shifted.
We compare relationship trends to counterfactual scenarios, in which countries collaborate with or cite other countries randomly. 
This exercise enables us to quantify the integration of third countries with Chinese and American AI research. 


Leveraging a large-scale and open-source publication dataset that contains all publications in AI and related fields, we generate a dynamic country-level collaboration, and citation network spanning the 1990-2023 period. 
Instead of calculating the shares of collaborations and citations among countries (like in \cite{silver2020us}) that is sensitive to scale, we quantify the significance of collaboration and citation links by comparing the observed weight to realizations of a network null model based on the maximum entropy principle. 
This approach allows us to investigate the significance of cross-country links in the two dimensions of collaborations and knowledge exchanges, and quantify the significance trends of every country with the US and China, in both dimensions.
Finally, we cluster these two-dimensional trajectories to identify archetypes of long-term divergence versus integration that characterize the evolving structure of the global AI research system.


We find that the USA and China have been diverging from each other in both collaborations and knowledge exchanges since the 2000s, emerging as two major poles of the international AI research network. 
Further countries integrate into this polarized landscape in distinctive pathways.
Major AI producers, such as the United Kingdom and Germany, integrate with the US but not with China.
A broad group of countries across different development levels converges exclusively toward China. Although China still lags behind in geographical coverage of significant partners, this finding signals its growing influence over international AI research that can challenge the dominance of the US.
By integrating simultaneously with both China and the United States, several European countries and other advanced economies position themselves as bridges across the poles, potentially gaining strategic advantages from connections to both AI research superpowers.

In sum, we show that the global AI research system is polarizing into two spheres of influence that can threaten global AI regulation. Yet, a group of doubly integrated countries, many of them European, still bridges both poles. 
Whether this bridging capacity is leveraged or eroded may determine the future integration of international AI research and consequently, the success of regulation strategies.

\section*{Results}\label{sec:res}
\subsection*{Data}
We collected data on scientific publications from OpenAlex \cite{priem2022openalex,chawla2022massive}, an open-source bibliographic database covering over 200 million records with information on publication dates, author affiliations, citation links, and disciplinary classifications based on the Scopus ASJC \cite{scopusASJC}. 
We collected articles published between 1990 and 2021 associated with the subfield ``Artificial Intelligence'', as well as papers published until 2023 (not necessarily related to AI) that cite them.
For each of them, we gathered the list of coauthors together with their institutional affiliation and filtered out articles for which no geographical information could be retrieved.
The resulting dataset contained 1'992'760 papers classified as AI, plus 1'239'696 further papers not classified as AI, but citing the AI articles.

\begin{figure}[b!]
    \centering
    \includegraphics[
    width=0.9\linewidth,
    height=\textheight,
    keepaspectratio]{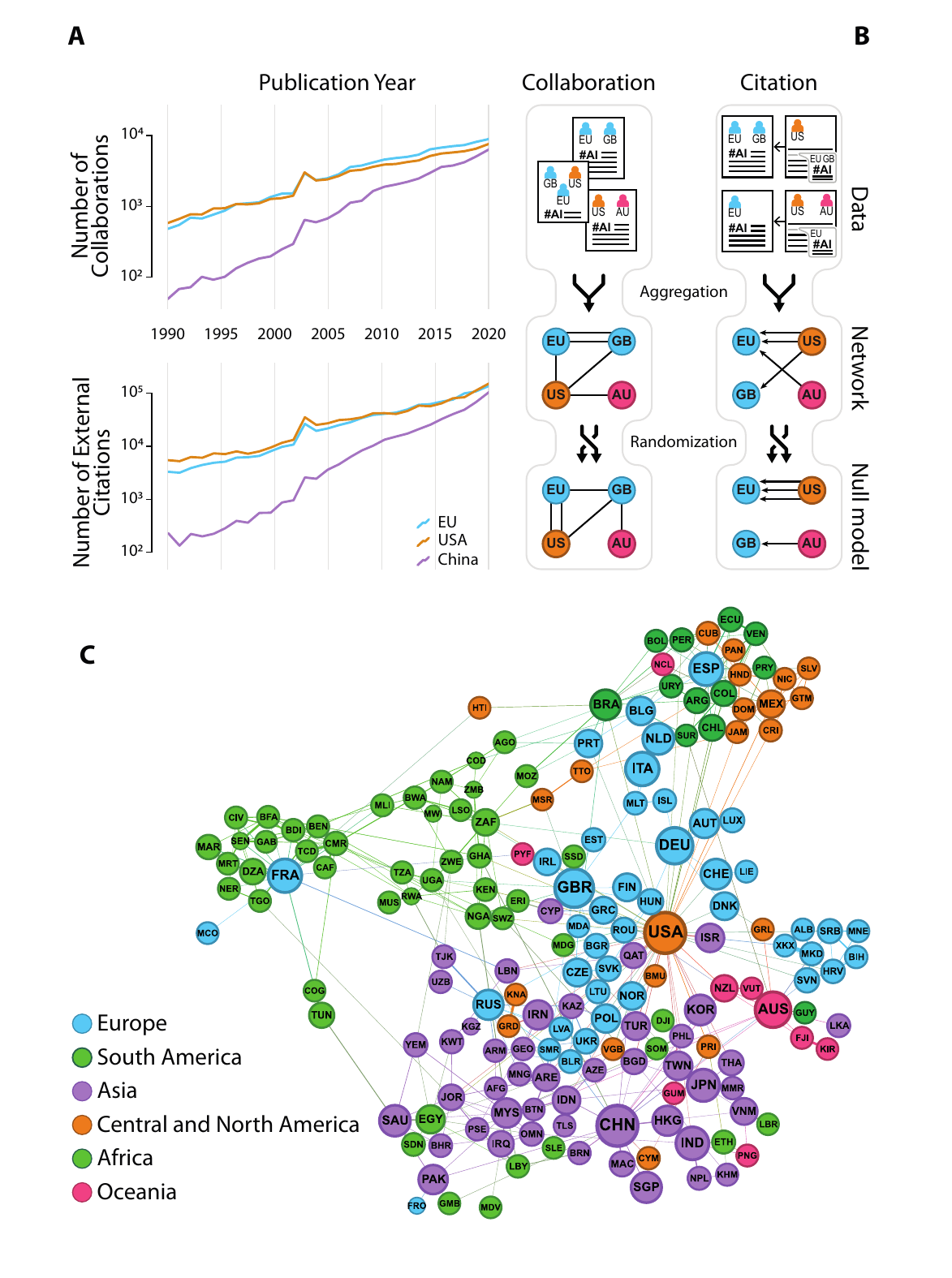}
    \caption{\textbf{Pipeline of the study}. 
    \textbf{A} Emergence of China as a global hub of AI research, measured by international collaborations and citations.
    \textbf{B} 
    We randomize temporal, weighted networks of international scientific collaborations and citations between countries to create synthetic networks that we for measuring the significance of observed cross-country links. 
    \textbf{C} The network of significant international collaborations for the period 2017-2021 illustrates that the US is linked with more countries across continents (46) than China (17) that is connected to neighbors and allies, while the EU-28 countries are only loosely connected to one another. Node size depicts number of papers, the edge significance threshold is two standard deviations difference between observed and randomized link weight.}
    \label{fig:fig1}
\end{figure}

Over the last thirty years, China has emerged as one of the world's scientific hubs.
While China was lagging behind the European Union (EU-28, including the United Kingdom) and the United States in the 90s, in terms of the number of international collaborations (i.e., papers whose authors are affiliated with institutions from China and at least one other country), and foreign citations, it now challenges their leadership and is likely to become the leading country in these figures (\cref{fig:fig1}\textbf{A}).
However, beyond the simple counting, it is crucial to consider which countries are collaborating with China, the US, and the European countries, and which are citing them.

To capture this structure, we modeled international collaborations in artificial intelligence as a temporal, undirected network (\cref{fig:fig1}\textbf{B}).
We built a network, i.e., a snapshot, for each publication year.
In each yearly snapshot, two countries are linked if they co-authored a paper (collaboration).
Each link is weighted by the number of collaborations.
Similarly, we modeled the citations among countries as a temporal directed network.
In each temporal snapshot, we put a link from country $i$ to country $j$ if a paper authored by country $i$ cites a paper authored by country $j$, weighting by the number of citations.

\subsection*{Significant patterns in collaboration and knowledge exchange networks}
The collaboration and citation networks provide insights into the relationships between countries in the production and exchange of knowledge on artificial intelligence.
Yet they do not provide information on how significant these relationships are relative to the countries' collaboration and citation volumes that impacts the weight of cross-country links and can distort their importance.
For instance, in 2021, Tunisia and France coauthored 52 papers, while Italy collaborated with France on 198 papers.
However, Tunisia collaborated on 352 papers in 2021, making those with France almost 15\% of its collaborations, whereas Italy's collaboration with France accounted for just 5\% of its 3,630 co-authorships.

We quantified the significance of the relationships by comparing the collaboration and citation networks to randomized versions leveraging a maximum-entropy approach \cite{mastrandrea2014enhanced}.
For each yearly temporal snapshot, we evaluated the expected link weights based on countries' total numbers of collaborations/citations and total number of connections, and compared them with the observed weights (\cref{fig:fig1}\textbf{B}, see Methods for details).
This null model allowed us to determine which relationships are stronger (or weaker) than we would expect based on the countries' collaboration and citation rates.

We explored present-day patterns of international collaborations by building a network of significant collaborations where we linked countries if, on average over the last five years (2017 to 2021), the z-score associated to their relationship was higher than 2, i.e., the observed number of collaborations was two standard deviations larger than that expected from the null model (\cref{fig:fig1}\textbf{C}).
Although China and the US have a similar number of international collaborations, their networks differ markedly. 



The US maintains significant ties with a diverse set of 46 countries across South America, Europe, Africa, and Asia. 
China, by contrast, has significant ties with only 17 countries, mostly geographically proximate or with strong historical connections, such as Pakistan, Vietnam, and Hong Kong. 
Notably, the collaboration between China and the US itself is not significant. 
Beyond these two poles, we identify tight-knit clusters of significant collaboration among South American countries, former Soviet republics, Balkan countries, and Arab countries. 
The EU-28, by contrast, forms a loosely connected, multi-core structure: 
the most productive members (e.g., the UK, Germany, Italy, and France) collaborate significantly with less productive neighbors but not with each other.
This suggests that the European AI research ecosystem lacks internal integration, despite the EU collectively having more international collaborations than both China and the US.

\subsection*{The dynamics of relationships' significance}
\begin{figure}[b!]
    \centering
    \includegraphics[width=\linewidth]{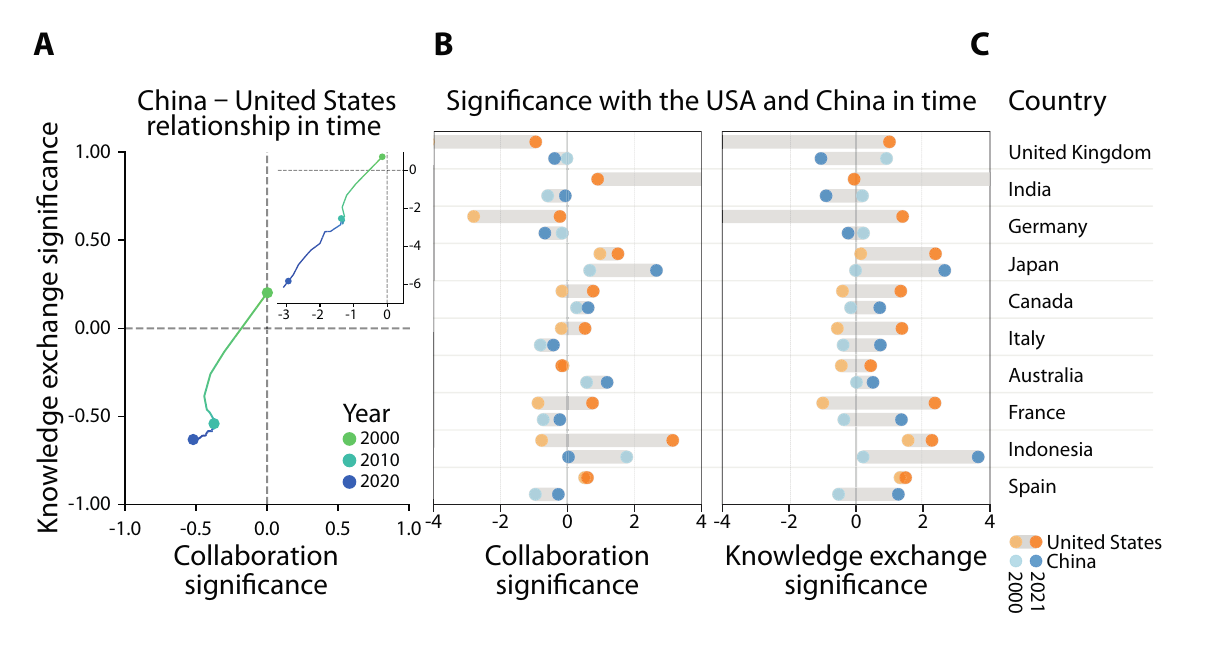}
    \caption{\textbf{Temporal evolution of the collaboration and knowledge exchange significance.}
    \textbf{A} Evolution of the China-US relationship between 2000 and 2021. 
    China and the US are collaborating and exchanging knowledge less than expected, and the significance of this gap has increased over time.
    Inset: Significance dynamics of the China-US relationship after controlling for the variance in the number of collaborations and knowledge exchanges intrinsic to the maximum-entropy approach (see Methods).
    \textbf{B-C} Variation in the relationships between the most prolific countries in AI with the US (orange) and China (blue), from 2000 to 2021.
    The significance of collaborations and knowledge exchanges has varied, with countries becoming more significantly related to the US (the UK and Germany), and others integrating with both (Japan, Canada, Italy, Australia, France, and Spain).
    The substantial change in India's relationship with the US may signal its emergence as a new global pole of AI research.}
    \label{fig:fig2}
\end{figure}

Beyond identifying the strongest connections between countries, a crucial aspect we investigated was how bilateral relationships had evolved over the past two decades.
We explored the evolution of the significance of the number of collaborations and of knowledge exchanges, i.e., the total number of citations from one country to the other and vice versa, by building a trajectory of z-scores over the years.  

Focusing on China and the US from 2000 to 2021, we found that the two countries have been diverging in both dimensions (\cref{fig:fig2}\textbf{A}). 
Although the raw volume of their collaborations and citations has increased (see Supplementary Information), this growth has consistently lagged behind what their respective research outputs would predict, and the gap is becoming increasingly significant over time.
This implies that, despite the potential benefits of cooperation \cite{alshebli2024china}, China and the USA are holding back on collaborations and knowledge exchanges.
The significance of this divergence gets stronger when we account for the variance in the number of collaborations and citations intrinsic to the maximum entropy model (inset of \cref{fig:fig2}\textbf{A}).

The China-US competition in AI research extends beyond their bilateral relationship, and it is also reflected in how they relate to other countries.
Therefore, we investigated how the ties of the most prolific countries in AI research with China and the US evolved between 2000 and 2021, finding markedly different trajectories (\cref{fig:fig2}\textbf{B}–\textbf{C}).

The UK and Germany became more significantly related with the US, both in collaborations and knowledge exchanges, while diverging from China. 
A second group, including Japan, Canada, Italy, Australia, France, and Spain, integrated with both China and the US.
Yet the magnitude of these shifts varies across countries.
For instance, the significance of collaboration and knowledge exchange between Japan and China has increased faster than that between Japan and the US, while we found the opposite for France.
Indonesia and India make other remarkable cases: 
the former intensified collaborations with the US while strengthening knowledge exchanges with China.
The latter dramatically degraded its relationship with the US, despite the two countries still significantly collaborating (see again \cref{fig:fig1}\textbf{C}).
Alongside its rapid growth in international collaborations and citations, this finding suggests that India may be emerging as a new global pole for AI research.  

\subsection*{Polarization and integration dynamics}

\begin{figure}[t!]
    \centering
    \includegraphics[
    width=0.92\linewidth,
    height=\textheight,
    keepaspectratio]{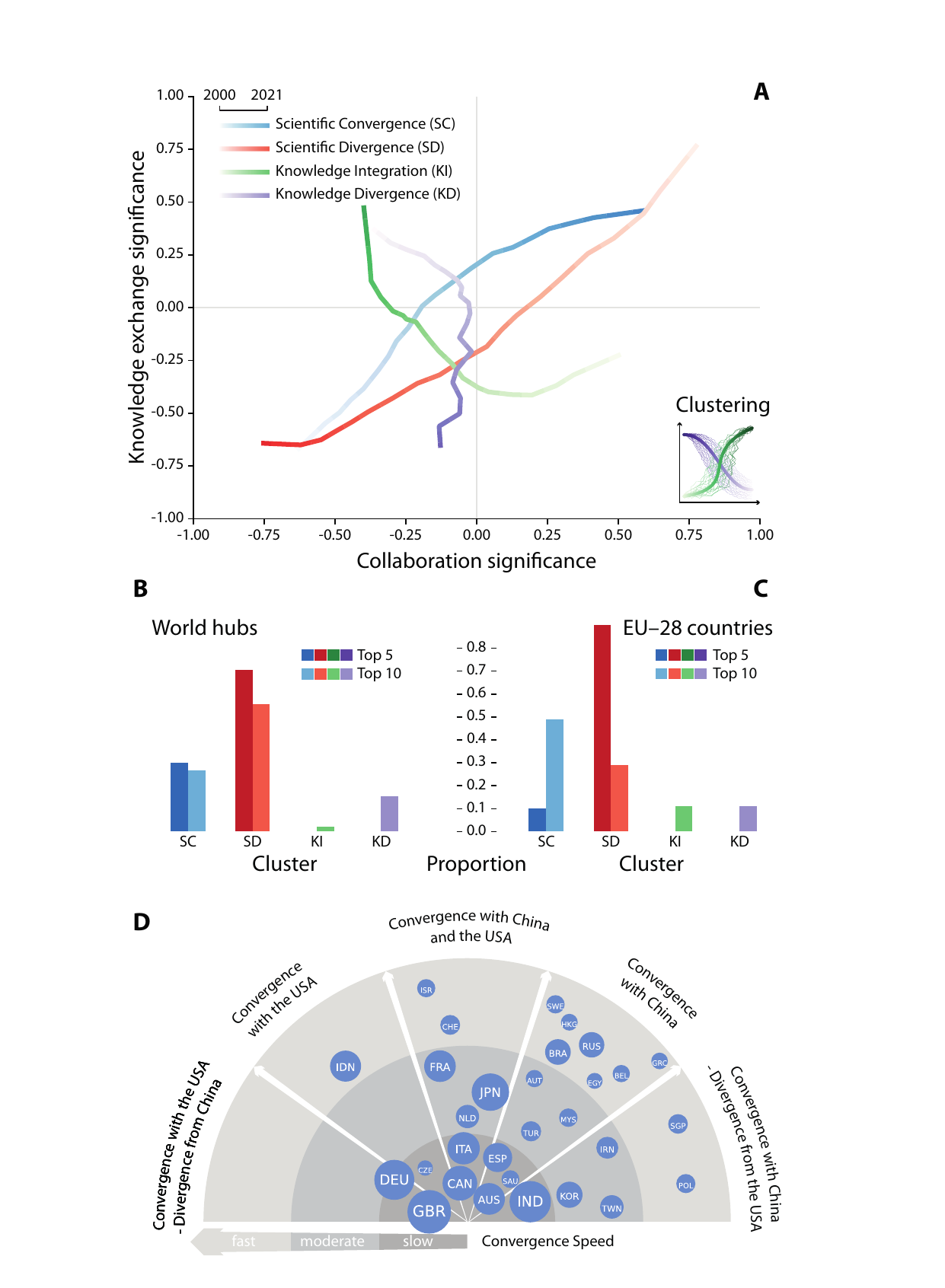}
    \caption{\textbf{Polarization and integration dynamics}. 
    \textbf{A} We detect four major trends in the dynamics of collaboration and knowledge exchange significance by clustering trajectories of country-to-country relationships. 
    \textbf{B-C} Distributions of links among global (\textbf{B}) and European (\textbf{C}) production hubs in the major trends suggest divergence of global AI hubs and a polarizing EU-28 network.
    \textbf{D} Trends of China's and the US' relationships with the 30 most productive countries in AI grouped by the direction and speed of trend with China and the US. Countries on the left versus on the right converge only with the US versus only with China, and countries in the middle converge with both poles.
    }
    \label{fig:fig3}
\end{figure}

The individual trajectories described above suggest recurring patterns in how bilateral relationships evolve over the decades.
Investigating if there are similar patterns in how these relationships develop is therefore essential for understanding the major tendencies of the global AI research ecosystem.
We inferred the main underlying trends by clustering significance trajectories using the k-means algorithm with Dynamic Time Warping (DTW) as the similarity metric (see Methods for details).
We identified four major trends (\cref{fig:fig3}\textbf{A}):
i) Scientific Convergence (SC), where the significance of both collaboration and knowledge exchange increases over time;
ii) Scientific Divergence (SD), where both significance scores decrease;
iii) Knowledge Integration (KI), where the knowledge exchange significance increases, with decreasing collaborations; and
iv) Knowledge Divergence (KD), where knowledge exchange significance decreases with stable collaborations.


We examined country-to-country relationships among the most productive countries, focusing on which of these four trends they follow.
We found that countries with the largest number of AI research papers are mostly diverging from one another (\cref{fig:fig3}\textbf{B}).
This suggests a generalized tendency toward global polarization in scientific research on AI.

A similar pattern holds within Europe. 
The most productive EU-28 countries are all diverging from each other, yet their relationships with less productive EU members tend toward scientific convergence (\cref{fig:fig3}\textbf{C}). 
This points to a European AI research network evolving into a hub-and-spoke structure: Major knowledge producers are becoming increasingly connected to smaller neighbors, while remaining poorly integrated with one another.

Finally, we examined how the 30 most productive AI countries align with China and the US over time, classifying them according to their convergence or divergence trajectories with both poles (see Methods; \cref{fig:fig3}\textbf{D}).

We found that the United Kingdom and Germany, the third- and fourth-most productive countries, respectively, are gradually converging toward the US while diverging from China (\cref{fig:fig3}\textbf{D}).
A large group, including Australia, Brazil, Russia, Spain, and several other developed and developing countries, is converging toward China. 
Among these, India, Poland, Singapore, South Korea, and Taiwan are simultaneously diverging from the US, providing the strongest evidence of active polarization and signaling China's growing centrality in the international AI research system. 
However, a group of countries resists this polarization: France, Italy, the Netherlands, Canada, Japan, and Israel are integrating with both China and the US, maintaining potentially advantageous connections to both knowledge ecosystems. 

We replicated these analyses using two alternative null models: A variance-corrected maximum-entropy approach and a Local Rewiring Algorithm \cite{maslov2002specificity,carstens2017switching}, that preserves node strengths in each network randomization, both of which corroborated our findings (see Supplementary Information).

\section*{Discussion}\label{sec:dis}


Our results highlight a core tension: a polarized AI race—resembling Cold War–style competition—undermines the prospects for ethical, broadly beneficial AI. 
Evidence that integration improves science \cite{schmallenbach2024global, alshebli2024china}, coupled with risks from fragmented regulation \cite{cihon2020fragmentation}, suggests that a purely bloc-based approach constrains diffusion and trust. 
Governance is pivotal: while the OECD anchors standard-setting \cite{schmitt2022mapping}, China is not a member; Europe’s attempts to cooperate with China in research funding have seen limited success, and access to close-to-market actions remains restricted.
The question is therefore not whether to integrate, but how to integrate under constraints.

Our network analysis reveals a long-run divergence between the United States and China in both collaboration and knowledge exchange, with each forming a pole of the global AI system. 
Yet the landscape is not binary: some countries integrate asymmetrically with one pole, while others become doubly integrated, deepening ties with both.
These bridge positions matter: they buffer polarization, sustain cross-bloc knowledge flows, and 
reduce the risk of regulatory fragmentation.

For both the United States and China, our results imply a twin challenge in “winning” the international AI race: polarization is pulling them into rival poles precisely when the most productive science arises in integrated networks.
Success, therefore, hinges less on headline capabilities than on retaining and convincing the bridge countries that connect both sides.
This requires limiting standards fragmentation through interoperable rules and evaluations. 
It demands calibrating research-security measures with pre-competitive carve-outs that keep collaboration viable, and investing in open benchmarks, data, and causal evaluation to demonstrate real benefits to partners.
The European Union, whose member states span the different integration pathways, is particularly well positioned to play this bridging role. 
At the same time, credible guardrails for the dark side—misuse, surveillance, unequal access—are essential to sustain legitimacy and trust; without them, potential allies drift, and influence over the trajectory of global AI diminishes.

\section*{Methods}\label{sec:meth}
\subsection*{Null model of weighted networks}
We quantified the significance of the relationships between countries by comparing the observed evolution of the collaboration and knowledge exchange networks to randomized versions.
We employed a maximum-entropy approach—the Enhanced Configuration Model (ECM)
\cite{mastrandrea2014enhanced}, which constructs a set of weighted networks that preserve, on average, the same degree and strength distributions as the original networks while randomizing other structural features. 
This approach allows us to identify relevant network features that cannot be simply explained by countries' collaboration and citation rates alone \cite{squartini2011analytical,cimini2019statistical}.
In both networks, node degree represents the number of countries with which a state collaborates. 
Node strength, however, differs between the two: in the collaboration network, it represents the number of papers a country co-authors, while in the knowledge exchange network, it corresponds to the total citations received and given by a country.

The maximum-entropy approach aims at finding the probability distribution $P(\mathbf{W})$ over the space of weighted networks $\mathbf{W}$ that 1) maximizes the randomness of the ensemble, while 2) the expected degree and strength sequences are constrained (on average) to certain values $\vec{k}_C$ and $\vec{s}_C$. 
This is realized by imposing that $P(\mathbf{W})$ maximizes the Shannon's entropy:
\begin{equation}
    S = -\sum_\mathbf{W}P(\mathbf{W})\,\mathrm{ln}\,P(\mathbf{W}),
\end{equation}
subject to constraints 
\begin{equation}
    \sum_\mathbf{W}\vec{k}_\mathbf{W}P(\mathbf{W}) = \vec{k}_C, \quad \sum_\mathbf{W}\vec{s}_\mathbf{W}P(\mathbf{W}) = \vec{s}_C.
\end{equation}
The constrained maximization can be solved \cite{garlaschelli2009generalized} by introducing two sets of Lagrange multipliers, $\vec{x}$ (controlling for the degrees) and $\vec{y}$ (controlling for the strengths), which allows us to write the (conditioned) probability distribution over the ensemble of weighted networks as
\begin{equation}
    P(\mathbf{W}|\vec{x},\vec{y}) = \prod_{i<j}q_{ij}(w_{ij}|\vec{x},\vec{y}),
\end{equation}
where 
\begin{equation}
    q_{ij}(w|\vec{x},\vec{y}) = \frac{(x_i x_j)^{\theta(w)}(y_iy_j)^w(1-y_iy_j)}{1-y_iy_j+x_ix_jy_iy_j}
\end{equation}
is the probability that a link between nodes $i$ and $j$ exists and have weight $w$.
Here $\theta(w)$ denotes the Heaviside function, i.e., $\theta(w)=1$ if $w>0$, and zero otherwise.

The distribution $P(\mathbf{W}|\vec{x},\vec{y})$ is a function of the free parameters $\vec{x}$ and $\vec{y}$.
To use the graph ensemble as a null model for an empirical network, say $\mathbf{W}^*$, we need to determine them uniquely and unbiasedly.
We can achieve this by finding $\vec{x}^*$ and $\vec{y}^*$ that maximizes the (log-)likelihood of generating $\mathbf{W}^*$ \cite{garlaschelli2008maximum}, formally
\begin{equation}
    \vec{x}^*,\vec{y}^* = \underset{\vec{x},\vec{y}}{\max}\, \mathcal{L}(\vec{x},\vec{y}) = \underset{\vec{x},\vec{y}}{\max}\,\mathrm{ln}\,P(\mathbf{W}|\vec{x},\vec{y})
\end{equation}
Previous research \cite{mastrandrea2014enhanced} has shown that $\vec{x}^*$ and $\vec{y}^*$ are those that solve the system of equations
\begin{align}
    \sum_{i<j}\frac{x_ix_jy_iy_j}{1-y_iy_j+x_ix_jy_iy_j} = k_i^* \\
    \sum_{i<j}\frac{x_ix_jy_iy_j}{(1-y_iy_j)(1-y_iy_j+x_ix_jy_iy_j)} = s_i^*
\end{align}
where $\vec{k}^*$ and $\vec{s}^*$ are the degree and strength sequences of the weighted network under study, $\mathbf{W}^*$.
Finally, we can sample randomized versions from the distribution $P(\mathbf{W}|\vec{x}^*,\vec{y}^*)$ \cite{squartini2015unbiased}.

Since the networks evolve over time, we evaluate the probability distribution for each publication year, generating 500 randomized networks per year.
We derived the expected numbers of collaborations and citations for each country-to-country relationship by averaging over the set of randomized versions.
Finally, we measured the significance of these relationships as the number of standard deviations $z$ by which the observed collaborations or knowledge exchanges, $w_{ij}^*$, differ from their expected values in the null models, namely
\begin{equation}
    z = \frac{w_{ij}^* - \mathrm{E}[w_{ij}]}{\sigma[w_{ij}]}.
\end{equation}

\subsection*{Clustering trajectories in the z-score plane}
We inferred global trends in country relationships by clustering the trajectories in the collaboration and knowledge exchange significance plane.
We analyzed the period between 2000 and 2021, retaining only trajectories with finite values of the z-scores over the entire time span.

We first smoothed the time series by applying the Savitzky–Golay filter \cite{savitzky1964smoothing}, fitting subsequent windows of 5 publication years with a 1st-order polynomial.
This isolated the long-term trend in the trajectory, while filtering out short-term noise.
We then rescaled the trajectories so that they span the interval [-1,1] in each dimension.
This normalization allowed us to focus on the shape of the trajectories—the direction along which they evolve in the z-score plane—rather than their absolute values.

We partitioned the trajectories into 4 clusters with the k-means algorithm \cite{lloyd1982least} using Dynamic Time Warping (DTW) \cite{sakoe2003dynamic} as the distance metric between the time series.
The cluster centroids represent average trajectories in the z-score plane, capturing the dominant global trends in collaboration and knowledge exchange significance dynamics.

\subsection*{Integration dynamics with China and the US}
We analyzed integration dynamics with China and the US, considering only those countries whose relationships with the two hubs fall into the Scientific Convergence or Scientific Divergence clusters.  
We partitioned the 30 most productive countries in AI research into five groups based on the dynamics of their relationship with China and the US.
Two groups comprise countries whose relationship with China (the US, respectively) is in the Scientific Convergence cluster, while the relationship with the US (China, respectively) is in the Scientific Divergence cluster.
Two other groups contain countries converging with China or the US, without a clear convergence/divergence relationship with the other country.
Countries converging with both China and the US make up the last group.

We measured the convergence speed as the magnitude of the variation in the collaboration and knowledge exchange significance plane over the 2000-2021 period.
Considering pairs of countries that are converging with each other, we denote the significance of their collaborations and knowledge exchanges at a given publication year $y$ as $z_{\mathrm{col}}(y)$ and $z_{\mathrm{exc}}(y)$, respectively.
Formally, we define their convergence speed as 
\begin{equation}
    v = \sqrt{\left[z_{\mathrm{col}}(2021) - z_{\mathrm{col}}(2000)\right]^2 + \left[z_{\mathrm{exc}}(2021) - z_{\mathrm{exc}}(2000)\right]^2}.
\end{equation}

We further split countries within each group into three subgroups based on their convergence speed with China and/or the US.
We considered countries with a convergence speed $v<2$ to be slowly converging, those with speed $2\leq v < 3$ to be moderately-fast converging, while those with speed $v \geq 3$ to be fast converging.
For countries converging with both China and the US, we considered the largest speed and split the group accordingly.

\backmatter

\bmhead{Acknowledgements} 
L.G. thanks Claudia Acciai, Roberta Sinatra, Lingfei Wu, and the ANETI Lab team for their insightful suggestions. 
L.G. acknowledges support from the Villum Foundation (project no. 57396) at the University of Copenhagen.
R.D.C. acknowledges support from the Lagrange Project of the ISI Foundation funded by CRT Foundation. B.L. acknowledges support from the Momentum Grant of the Hungarian Academy of Sciences.

\bmhead{Data availability} The data from OpenAlex used in this study can be downloaded using a dedicated API \url{https://docs.openalex.org/how-to-use-the-api/api-overview}.

\bmhead{Code availability} The code developed in this study will be made available at \url{https://github.com/lgajo/polarization-integration-AI.git} upon publication.

\bibliography{sn-bibliography}

\end{document}


\title[Article Title]{Supplementary Information of the manuscript ``Polarization and Integration in Global AI Research''}


\author{\fnm{Luca} \sur{Gallo}}

\author{\fnm{Riccardo} \sur{Di Clemente}}

\author{\fnm{Balázs} \sur{Lengyel}}

\maketitle

\section*{China-US collaborations and knowledge exchanges over time}
\begin{figure}[h!]
    \centering
    \includegraphics[width=0.625\linewidth]{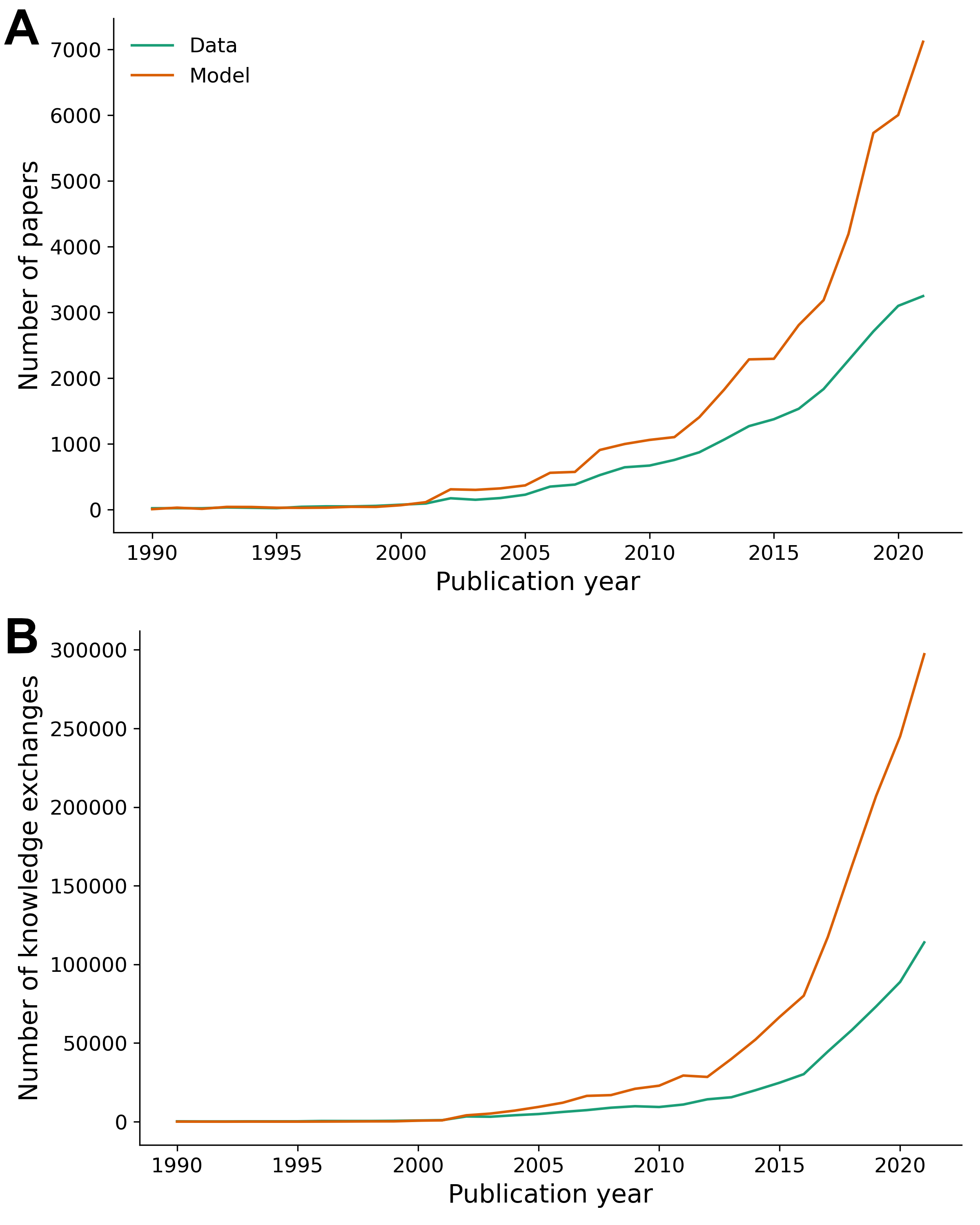}
    \caption{\textbf{Expected and observed number of collaborations and knowledge exchanges}. 
    \textbf{A} Evolution of the number of collaborations between China and the US from 1990 to 2021. 
    The number of collaborations has increased slower than expected from the production rates of the two countries.
    \textbf{B} Evolution of the number of knowledge exchanges between China and the US from 1990 to 2021. 
    The number of knowledge exchanges has increased slower than expected from the countries' citation rates.
    }
    \label{fig:figS0}
\end{figure}

In the main text, we discussed how China and the United States are collaborating and exchanging knowledge less than expected from their production and citation rates.
This, however, does not imply that (i) China and the US co-author few articles or cite each other rarely, and (ii) the number of collaborations and citations between the two countries is increasing over time.
China and the US, in effects, are the countries with the largest number of collaborations and knowledge exchanges since 2008, and these numbers have steadily increased since the 90s (\cref{fig:figS0}\textbf{A-B}).
However, the number of collaborations and knowledge exchanges expected from their production and citation rates is larger and has increased faster (\cref{fig:figS0}\textbf{A-B}).

\section*{Controlling for strength variance in the maximum-entropy null model}
In the main text, we showed how we quantified the significance of country-to-country relationships using a maximum-entropy approach named Enhanced Configuration Model (ECM) \cite{mastrandrea2014enhanced}.
The ECM is a null model that generates random versions of a given network with the same degree and strength sequences.
However, the model preserves these characteristics only on average across the ensemble of random networks, not in each individual realization.
Therefore, the degree and strength sequences vary across different random versions, which in turn substantially influences the link weights.
In particular, we expect the variance of the link weights in the ECM to be larger than in a null model where the constraint on the degree and strength is respected in each realization, i.e., in a microcanonical framework \cite{cimini2019statistical} (see also the following section of this SI).
This, in turn, would lead to generally smaller values of the statistical significance.

We investigated this effect by measuring the significance of country-to-country relationships after controlling for the variance of degrees and strengths over different randomized versions.
In particular, we performed a linear regression analysis. 
For each link $(i,j)$ in the $r$-th random network, we modeled its weight, $w_{ij,r}$, as
\begin{equation}
    w_{ij,r} = \beta_{ij,0} + \beta_{ij,1} s_{i,r} + \beta_{ij,2} s_{j,r} + \beta_{ij,3}\sqrt{s_{i,r}s_{j,r}} + \varepsilon_{ij,r}, \quad \mathrm{with} \; r\in{1,\dots,500}
\end{equation}
where $s_{i,r}$ and $s_{j,r}$ are the strengths of nodes $i$ and $j$, respectively.
We excluded node degrees from the model as they exhibited no correlation with link weights. 
The best-fit regression residuals quantify the amount of weight that cannot be explained by the node strengths, which is what we were interested in quantifying.
Therefore, we can evaluate the significance of country-to-country relationships as the number of standard deviations by which the residual weight of the empirical links (after controlling for the node strengths) deviates from the expected residuals.    

\begin{figure}[t!]
    \centering
    \includegraphics[width=\linewidth]{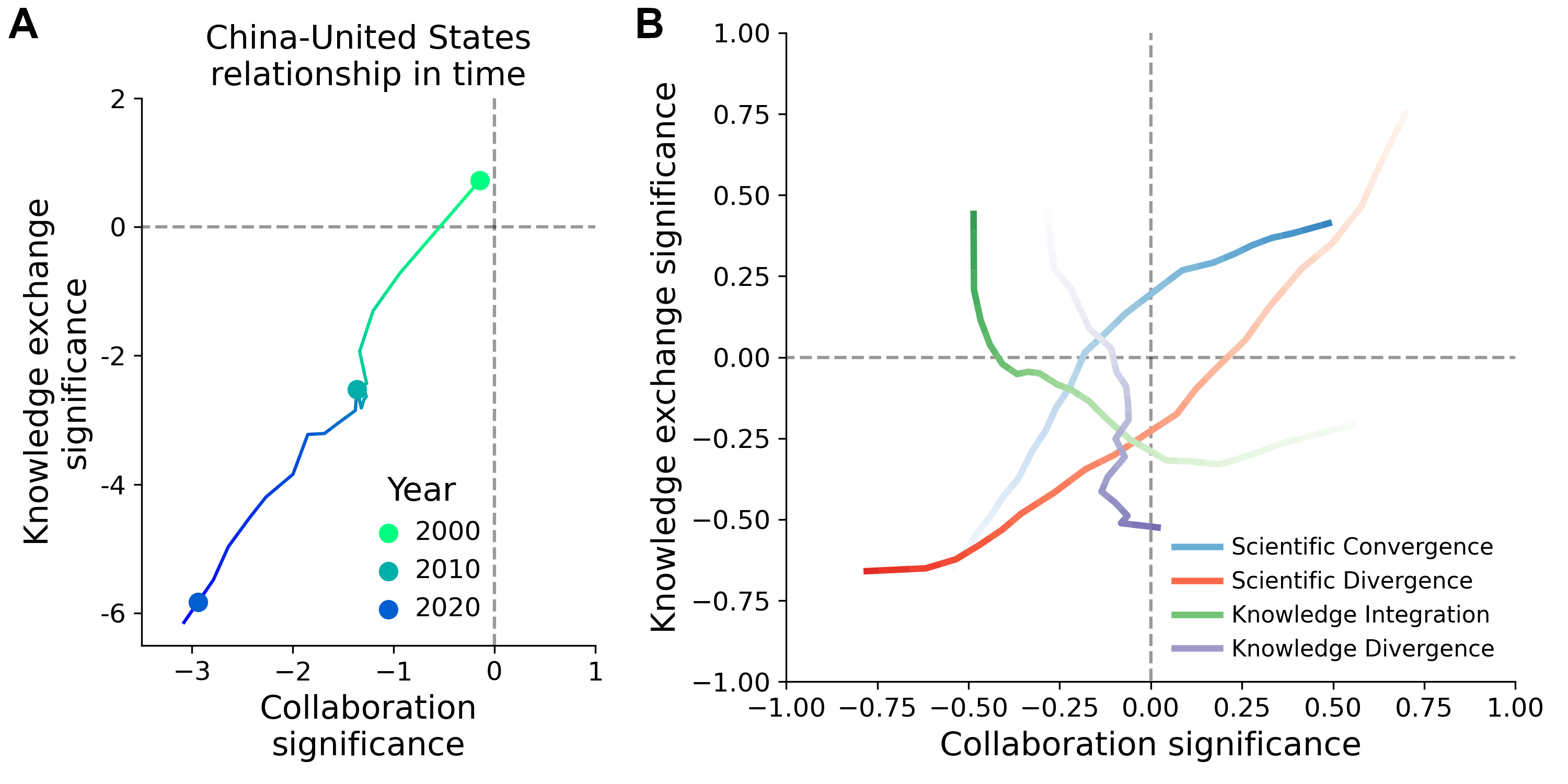}
    \caption{\textbf{Trends in the z-score dynamics using the modified maximum-entropy approach}. 
    \textbf{A} Evolution of the China-US relationship between 2000 and 2021. 
    China and the US collaborate and exchange knowledge less than expected, with the gap in significance increasing over time.
    \textbf{B} Trends in the dynamics of collaboration and knowledge exchange significance. 
    We identified pairs of countries that are integrating (Scientific Convergence) or diverging (Scientific Divergence) in both collaborations and knowledge exchanges, and pairs mostly increasing (Knowledge Integration) or decreasing (Knowledge Divergence) the significance of their knowledge exchanges.
    }
    \label{fig:figS1}
\end{figure}

We investigated the relationship between China and the US from 2000 to 2021 using this modified approach, finding consistent results: China and the US are diverging in both collaborations and knowledge exchanges (\cref{fig:figS1}\textbf{A}).

We inferred the major global trends in collaboration and knowledge-exchange dynamics by clustering trajectories in the significance plane.
The identified clusters similar to those described in the main text(\cref{fig:figS1}\textbf{B}):
i) Scientific Convergence (SC), with both collaboration and knowledge exchange significance increasing over time;
ii) Scientific Divergence (SD), with both decreasing;
iii) Knowledge Integration (KI), where the knowledge exchange significance increases; and
iv) Knowledge Divergence (KD), where knowledge exchange significance decreases.

\begin{figure}[t!]
    \centering
    \includegraphics[width=\linewidth]{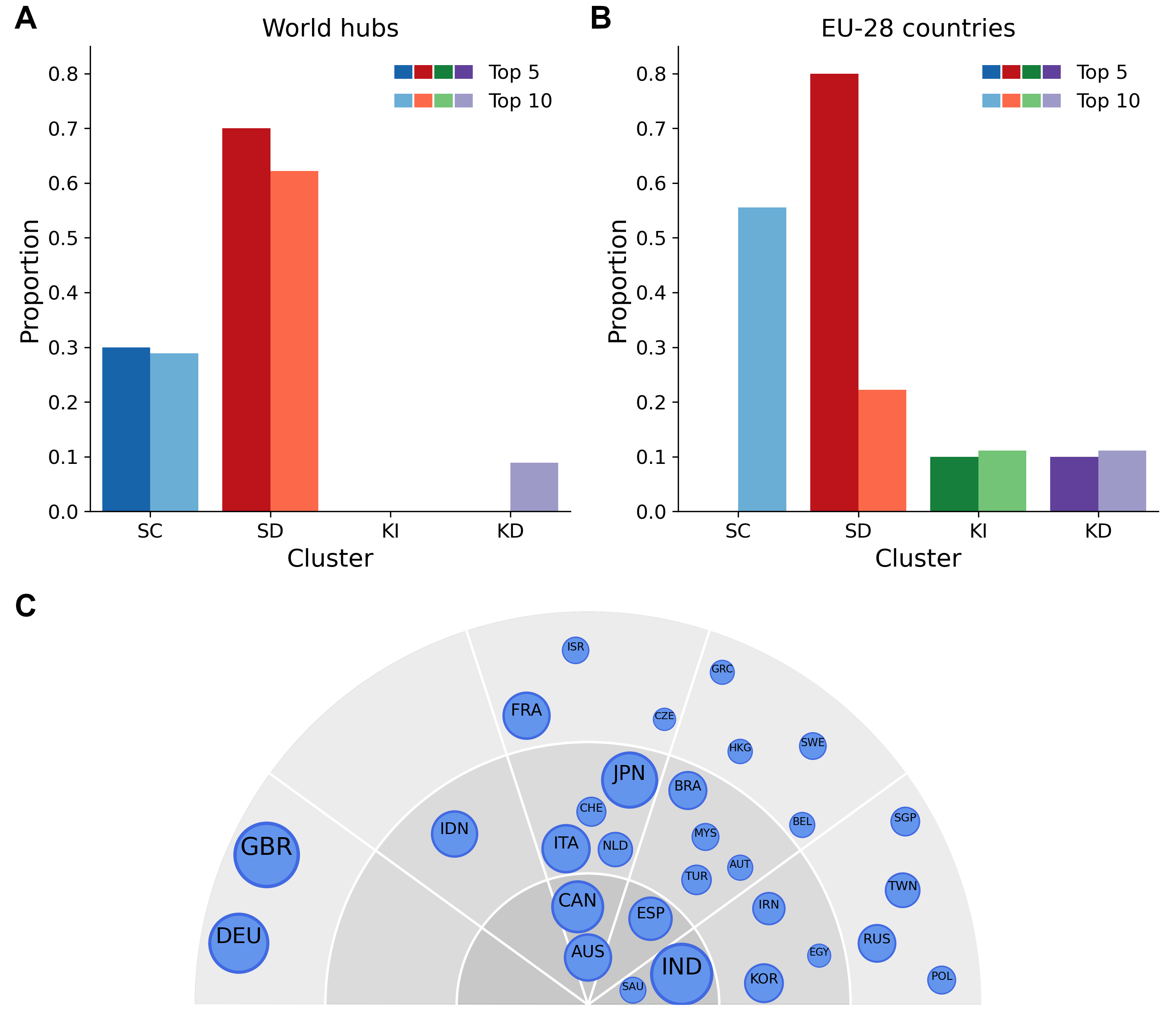}
    \caption{\textbf{Polarization and integration dynamics using the modified maximum-entropy approach}. 
    \textbf{A-B} Distributions of country-to-country relationships among global (\textbf{A}) and European (\textbf{B}) production hubs in the major dynamical trends.
    In both panels, darker colors refer to relationships among the 5 most productive countries, and lighter colors among the 10 most productive countries.
    Global hubs are diverging, while the EU-28 is shifting towards a core-periphery structure with divergent hubs.
    \textbf{C} Trends of China's and the US' relationships with the 30 most productive countries in AI.
    We grouped countries based on whether they converge with or diverge from China and the US.
    Germany and the United Kingdom are converging rapidly with the US while diverging from China. 
    Key AI producers are converging on China, with some simultaneously diverging from the US (e.g., India, Singapore, South Korea, Taiwan). 
    Important players (e.g., Canada, France, Italy, Japan) are integrating with both China and the US.
    }
    \label{fig:figS2}
\end{figure}

We analyzed which country-to-country relationships follow these four trends.
In agreement with the results shown in the main text, we found that the most productive countries in AI research are diverging (\cref{fig:figS2}\textbf{A}).
We found the same pattern when looking at EU-28 production hubs.
Yet, we observed a trend toward scientific convergence when including less productive countries (\cref{fig:figS2}\textbf{B}), confirming the robustness of the dynamics explained in the main text.

Finally, we once again explored the relationships of the 30 most productive countries with China and the US.
We grouped countries based on how they converge to or diverge from the two main players in AI research.
Here, we considered slow convergence speeds $v < 2$, moderately-fast $2 \leq v < 4$, and fast $v \geq 4$. 
We found results similar to those reported in the main text (\cref{fig:figS2}\textbf{C}). 
Germany and the United Kingdom are converging toward the US while diverging from China, though we found the convergence to be fast.
Key countries, such as Brazil, Russia, and Spain, as well as smaller countries like Belgium, Sweden, Malaysia, and Saudi Arabia, are converging on China.
India, Poland, Singapore, South Korea, and Taiwan are also diverging from the US.
Various European countries, such as France, Italy, and the Netherlands, as well as Canada, Japan, and Israel, are integrating with both China and the US.
We also observed some differences compared to the approach adopted in the main analysis:
For instance, we found that Australia was converging not only with China but also with the US.
Moreover, we found that Russia and Saudi Arabia are diverging from the US, a finding not present in the main analysis.
Overall, our modified maximum-entropy approach corroborates the key results shown in the main text. 

\section*{Randomization of the networks using the local rewiring algorithm}
In the main text, we showed how we used a maximum-entropy approach to evaluate the expected numbers of collaborations and knowledge exchange between two countries. 
We corroborated the results obtained with this method by using an alternative null model of our networks.
In particular, we generated random versions of the collaboration and citations networks using a version of the local rewiring algorithm (LRA) introduced in \cite{maslov2002specificity}, adapted to the case of weighted networks \cite{carstens2017switching}.
This approach makes a valuable alternative, as it preserves the strengths of the nodes in each individual realization, compared to the approach we employed, which preserves the strength sequence only on average across the ensemble or random versions.

Both the collaboration and the citation networks can be modeled as a multigraph, namely a graph where two nodes can be connected by more than one link.
In our context, the number of links corresponds to the number of collaborations or citations in a given year.
The citation network is a directed multigraph, where links from A to B are not the same as links from B to A. 
For this network, we performed the following rewiring algorithm:
We randomly select a pair of directed links, say $\mathrm{A}\rightarrow \mathrm{B}$ and $\mathrm{C}\rightarrow \mathrm{D}$, and rewire them to generate two new links, $\mathrm{A}\rightarrow \mathrm{D}$ and $\mathrm{C}\rightarrow \mathrm{B}$.
This swapping preserves both the in-strengths and out-strengths of each nodes, corresponding to the total number of citations given and received. 
The rewiring is accepted with a probability $p=1/w_{\mathrm{A}\rightarrow\mathrm{B}}w_{\mathrm{C}\rightarrow\mathrm{D}}$, where $w_{x\rightarrow y}$ denotes the number of citations from $x$ to $y$ before the rewiring. 
The acceptance probability guarantees an unbiased randomization \cite{carstens2017switching}:
Links with higher multiplicity are more likely to be selected, so they are more likely to be shuffled.
This in turn would make the random versions where links have, on average, a lower multiplicity more likely, thus biasing the expected number of citations towards lower values. 
The acceptance probability balances out the higher sampling probability, thus ensuring an unbiased rewiring process. 
Finally, if self-loops are generated, the rewiring is always rejected.
Repeating the rewiring process multiple times, one can thus generate a randomized version of the original network.
In our implementation, we attempted a number of swaps equal to a hundred times the number of links in the citation multigraph.

We followed a similar procedure for the collaboration network:
We select a pair of undirected links, $\mathrm{A}\text{---} \mathrm{B}$ and $\mathrm{C}\text{---} \mathrm{D}$, and rewire them to generate two new links, either $\mathrm{A}\text{---} \mathrm{D}$ and $\mathrm{C}\text{---} \mathrm{B}$, or $\mathrm{A}\text{---} \mathrm{C}$ and $\mathrm{B}\text{---} \mathrm{D}$, with probability 1/2.
The swap preserves the strength of the nodes, representing the total number of collaborations.
Again, we accept the rewiring with a probability $p=1/w_{\mathrm{A}\mathrm{B}}w_{\mathrm{C}\mathrm{D}}$, where $w_{xy}$ denotes the link multiplicity before the rewiring, i.e., the number of collaborations between $x$ and $y$, and we prevent the formation of self-loops.
We repeated the process multiple times to generate a randomized network.
In particular, we attempted to rewire a number of pairs equal to a hundred times the number of links in the collaboration multigraph.

For both networks and for each yearly snapshot, we generated 1000 randomized networks.
We constructed a new randomized version by reshuffling the last randomized version that we generated rather than the empirical network. 
This way, we reduced the statistical dependence deriving from seeding the algorithm on the same network, thus preventing potential biases coming from having explored the ensemble of random versions only close to the empirical network \cite{ansmann2011constrained}. 

As the collaboration and citations networks evolve in time, we performed the rewiring on each yearly snapshot separately.
Finally, we measured the significance of country-to-country relationships in each year has the number of standard deviations by which the multiplicity of a link deviates from its expected value, calculated as the average over the set of randomized versions.

\begin{figure}[t!]
    \centering
    \includegraphics[width=\linewidth]{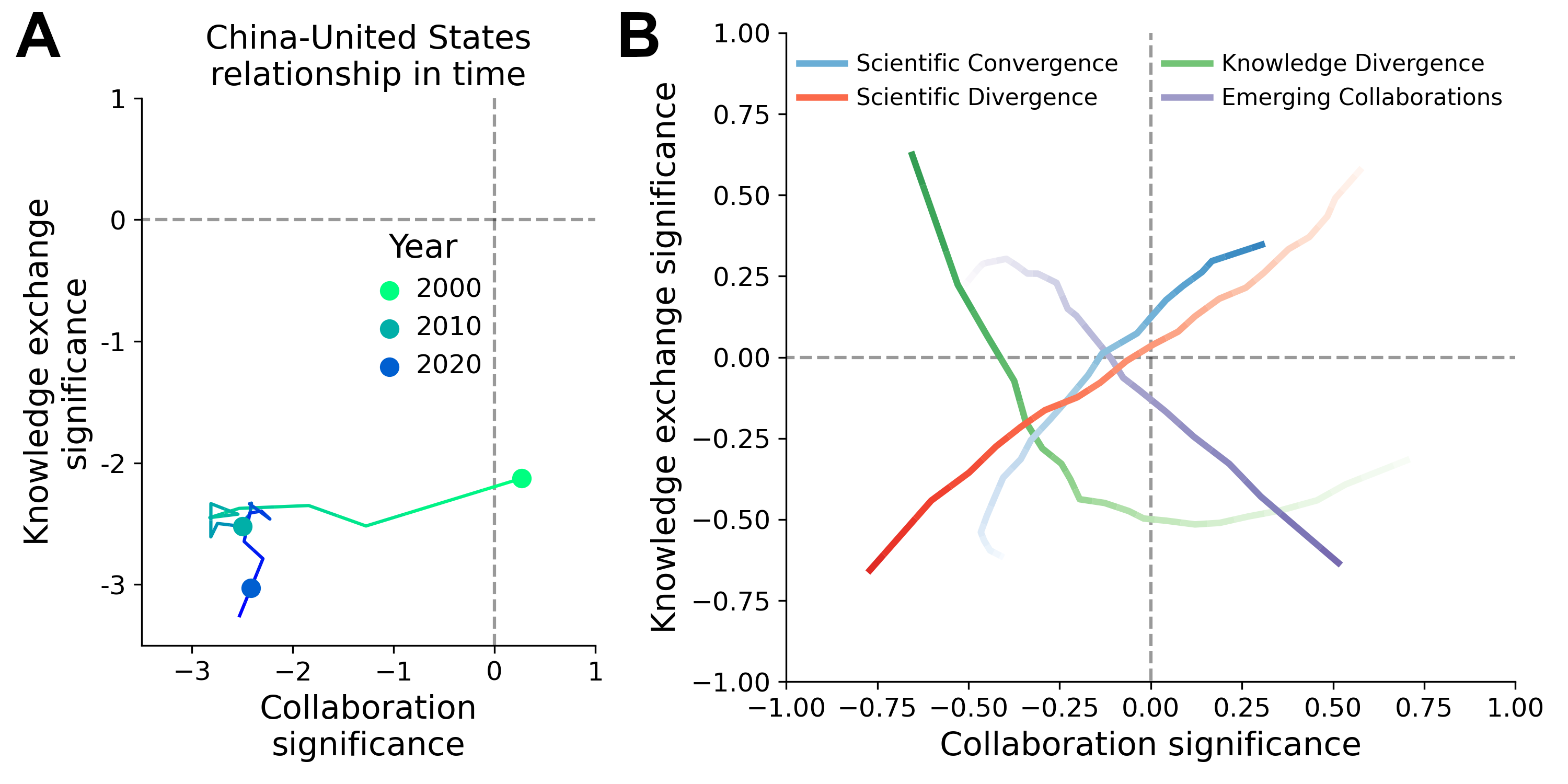}
    \caption{\textbf{Trends in the z-score dynamics using the local rewiring algorithm}. 
    \textbf{A} Evolution of the China-US relationship between 2000 and 2021. 
    China and the US collaborate and exchange knowledge less than expected, with the gap in significance increasing over time.
    \textbf{B} Trends in the dynamics of collaboration and knowledge exchange significance. 
    We identified pairs of countries that are integrating (Scientific Convergence) or diverging (Scientific Divergence) in both collaborations and knowledge exchanges, and pairs mostly increasing (Knowledge Integration) or decreasing (Knowledge Divergence) the significance of their knowledge exchanges.
    }
    \label{fig:figS3}
\end{figure}

We analyzed the relationship between China and the US from 2000 to 2021 using the LRA.
The result is qualitatively similar to that yielded by the maximum-entropy approach: China and the US are diverging in both collaborations and knowledge exchanges (\cref{fig:figS3}\textbf{A}).

We derived the global trends in collaboration and knowledge-exchange dynamics by clustering trajectories in the significance plane of country-to-country relationships.
We identified again four clusters(\cref{fig:figS3}\textbf{B}):
Three of them are in good agreement with what we obtained using the maximum-entropy approach:
i) Scientific Convergence (SC), with the significance of collaborations and knowledge exchanges increasing over time;
ii) Scientific Divergence (SD), with the significance decreasing for both;
iii) Knowledge Integration (KI), where the significance of knowledge exchanges increases.
We found a difference in the fourth cluster:
While the maximum-entropy approach leads to a Knowledge Divergence (KD) cluster, with the knowledge exchange significance decreasing and the collaboration significance slightly increasing, the LRA yields a cluster in which collaboration significance increases more than knowledge exchange significance.
We called this the Emerging Collaboration (EC) cluster.

\begin{figure}[t!]
    \centering
    \includegraphics[width=\linewidth]{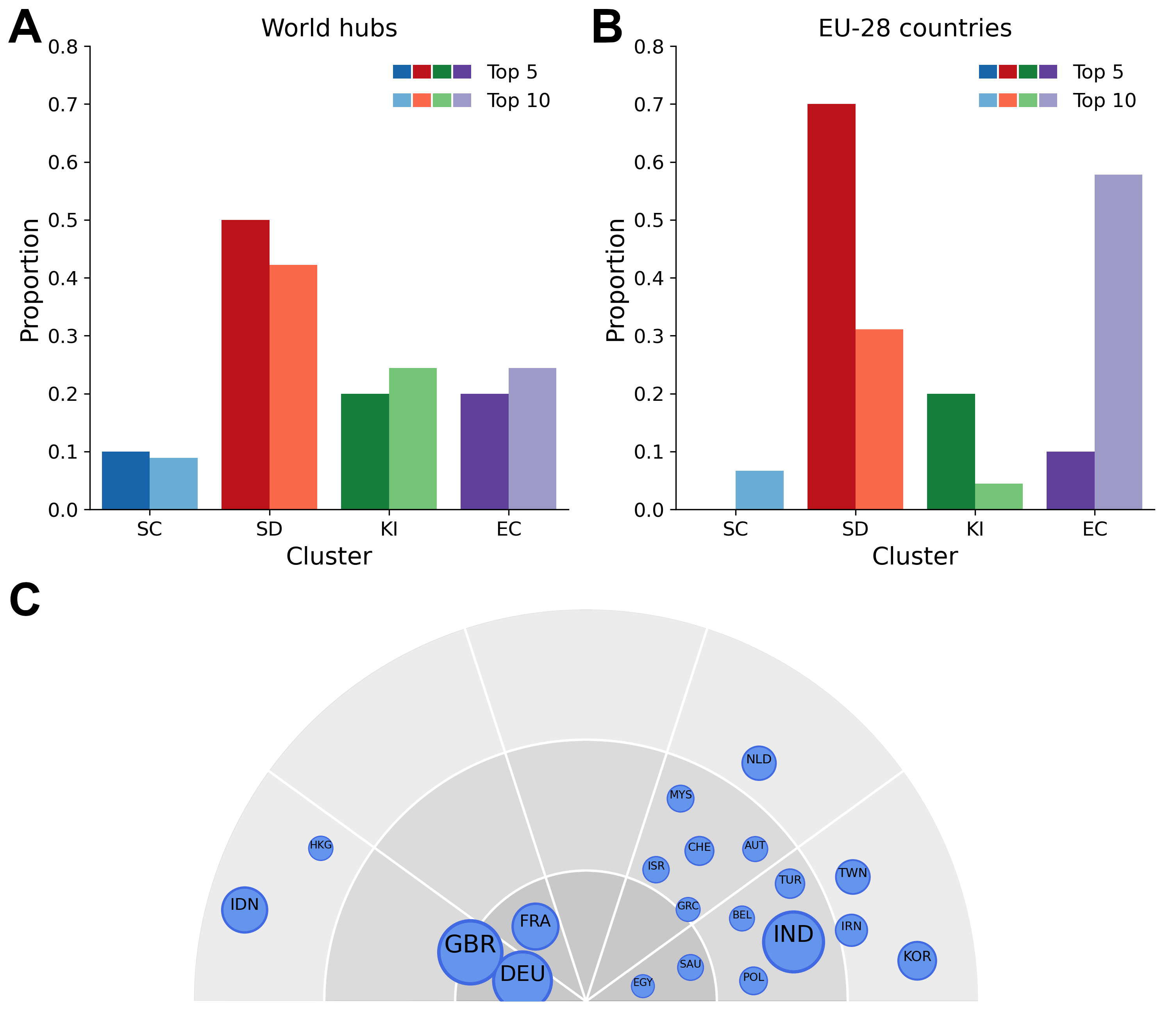}
    \caption{\textbf{Polarization and integration dynamics using the local rewiring algorithm}. 
    \textbf{A-B} Distributions of country-to-country relationships among global (\textbf{A}) and European (\textbf{B}) production hubs in the major dynamical trends.
    In both panels, darker colors refer to relationships among the 5 most productive countries, and lighter colors among the 10 most productive countries.
    Global hubs are diverging, while the EU-28 is shifting towards a core-periphery structure with divergent hubs.
    \textbf{C} Trends of China's and the US' relationships with the 30 most productive countries in AI.
    We grouped countries based on whether they converge with or diverge from China and the US.
    Germany and the United Kingdom are converging rapidly with the US while diverging from China. 
    Key AI producers are converging on China, with some simultaneously diverging from the US (e.g., India, Singapore, South Korea, Taiwan). 
    Important players (e.g., Canada, France, Italy, Japan) are integrating with both China and the US.
    }
    \label{fig:figS4}
\end{figure}

We explored country-to-country relationships within these four clusters.
We found that the most productive countries in AI are diverging, which aligns with the results obtained with the maximum-entropy approach (\cref{fig:figS4}\textbf{A}).
Similarly, the five most productive countries in the EU-28 are diverging.
Different from what we showed in the main text, however, we found that most country-to-country relationships fall into the Emerging Collaborations cluster when including less productive ER-28 countries (\cref{fig:figS4}\textbf{B}).
This is likely due to the significance of collaborations:
Among the trends inferred using the maximum-entropy approach, the Scientific Convergence cluster is the only one in which collaboration significance increases substantially. 
So, country-to-country relationships where the collaboration significance increases substantially are likely to end up in this cluster.
Using the LRA, instead, we obtained two clusters with increasing collaboration significance: the SC and the EC, with the latter showing a larger increase
Country-to-country relationships in which the significance of collaboration increases may thus end up in one of the two, depending on the dynamics of knowledge exchanges. 
This finding, therefore, agrees with the dynamics discussed in the main text, namely a scenario in which European hubs are diverging while integrating (at least in terms of collaborations) with smaller countries. 

Finally, we once again explored the relationships of the most productive countries with China and the US.
We again considered the 30 most productive countries in AI research, grouping them by how they converge with or diverge from China and the US.
Here, we considered countries to be diverging if their relationship fell in the Scientific Divergence cluster, and to be converging if it fell in the Scientific Convergence or the Emerging Collaborations clusters.
We considered convergence speeds to be slow if $v < 2$, moderately-fast if $2 \leq v < 3$, and fast if $v \geq 3$. 
Similarly to what we showed in the main text, Germany and the United Kingdom are slowly converging toward the US while diverging from China (\cref{fig:figS4}\textbf{C}).
Most countries, including Belgium, Malaysia, and Saudi Arabia, are converging on China, and some (India, Poland, South Korea, and Taiwan) are also diverging from the US.
However, we observed important differences compared to what we found with the maximum-entropy approach:
For instance, we found that import countries in AI research, like Australia, Canada, Italy, Japan, and Spain, are converging neither with China nor with the US.
Also, Hong Kong, which we found converging with China, appears to be converging with the US.
Crucially, we found no countries to be integrating with both China and the US:
Some of the countries we identified as bridging China and the US, such as France, Israel, the Netherlands, and Switzerland, are converging with only one of the two hubs. 

Overall, the LRA yields results that are in good agreement with the findings obtained using the maximum-entropy approach.
Yet the LRA faces a significant limitation: The number of (attempted) rewirings required to generate a randomized version of a network is not rigorously specified \cite{cimini2019statistical}.
If an insufficient number of rewirings is performed than the algorithm does not sample uniformly from the ensemble of networks, despite being theoretically unbiased.
This issue is particularly pronounced in weighted networks, where the probability of accepting the rewiring of a pair of links is inversely proportional to the product of the links' weights.
Therefore, as link weights in empirical networks are heavy-tailed distributed \cite{barthelemy2005characterization}, the acceptance probability of rewiring two heavy links can be close to zero, which in turn makes the required number of attempts computationally infeasible to achieve.
The maximum-entropy approach thus remains the most reliable methodology, as it guarantees an unbiased sample from the ensemble of randomized versions.

\bibliography{sn-bibliography}